\documentclass[aps,prl,superscriptaddress,twocolumn,floatfix,showpacs,citeautoscript,longbibliography]{revtex4-1}
\usepackage{amsmath,amsfonts,bm,mathbbol,graphicx,verbatim,mathrsfs,units,color}

\newcommand{\h}[1]{\hat{#1}}
\newcommand{\1}{\mathbb{1}}

\begin{document}

\title{The antiferromagnetic phase of the Floquet-driven Hubbard model}
\author{Nicklas Walldorf}
\affiliation{Center for Nanostructured Graphene (CNG), Department of Physics, Technical University of Denmark, DK-2800 Kongens Lyngby, Denmark}
\author{Dante M. Kennes}
\affiliation{Dahlem Center for Complex Quantum Systems and Fachbereich Physik, Freie Universit{\"a}t Berlin, 14195 Berlin, Germany}
\author{Jens Paaske}
\affiliation{Center for Quantum Devices, Niels Bohr Institute, University of Copenhagen, 2100 Copenhagen, Denmark}
\author{Andrew J. Millis}
\affiliation{Department of Physics, Columbia University, 538 West 120th Street, New Yok, NY 10027 USA}
\affiliation{Center for Computational Quantum Physics, Flatiron Institute, 162 5th Avenue, New York, NY 10010 USA}

\date{\today}

\begin{abstract}
A saddle point plus fluctuations analysis of the periodically driven half-filled two-dimensional Hubbard model is performed. For drive frequencies below the equilibrium gap, we find discontinuous transitions to time-dependent solutions. A highly excited, generically non-thermal distribution of magnons occurs even for drive frequencies far above the gap. Above a critical drive amplitude, the low-energy magnon distribution diverges as the frequency tends to zero and antiferromagnetism is destroyed, revealing the generic importance of collective mode excitations arising from a non-equilibrium drive.
\end{abstract}

\maketitle

The rapid development of stable, high-intensity, radiation sources has opened up new experimental horizons for  non-equilibrium control of material properties by application of tailored radiation fields~\cite{Mankowsky:Nonequilibrium,Basov:Towards,Tokura:Emergent}. An applied radiation field affects a material in two fundamentally different ways: by changing the Hamiltonian, and creating excitations. The former, commonly referred to as ``Floquet engineering'', offers an exciting route towards engineering new phases of driven matter~\cite{Rudner2013Jul,Bukov:Universal,Singla:THz,Mentink:Ultrafast,Claassen:All, Knap2016,Abanin:Effective,Kennes:Transient,Sentef:Light,Murakami2017,
Coulthard:Enhancement,Kitamura:Probing,Kennes:Floqueteng,
Tancogne:Ultrafast}. In some cases (e.g. integrable models with collisionless dynamics) mode excitation can lead to novel dynamical phases~\cite{Collado:Population}. However, in the generic situation, if too many excitations are created, the interesting phases can be destabilized~\cite{Peronaci:Resonant,Mitra:Nonequilibrium,Mitra:Current,Jose:Ultra}. The general consensus in the field has been that if the drive frequency is sufficiently detuned from the electronic transition energies, excitations may be neglected, allowing a focus on  ``Floquet engineering" aspects.

In this paper we investigate the physics of ac driven systems with drive frequency detuned from electronic transitions  via a theoretical study of the properties of the Hubbard model. This model is one of the paradigmatic systems of theoretical condensed matter physics, capturing the essential physics of electronic ordering and collective modes.  We focus on the effect of the ac drive on the antiferromagnetic phase and the associated collective modes. We find that even in the `detuned case', in which the ac drive does not produce a significant density of quasiparticle excitations, a highly non-equilibrium collective mode distribution is produced, with a remarkable dependence on the drive amplitude suggestive of a dynamical quantum phase transition. Above a critical drive amplitude, the non-equilibrium distribution of collective modes leads to a destruction of long-ranged antiferromagnetic order, possibly even for dimensions higher than two. 
These findings suggest that collective mode distribution effects may be important more broadly in the physics of Floquet-driven phases.
\begin{figure}[b]
  \centering
  \includegraphics[width=0.95\columnwidth]{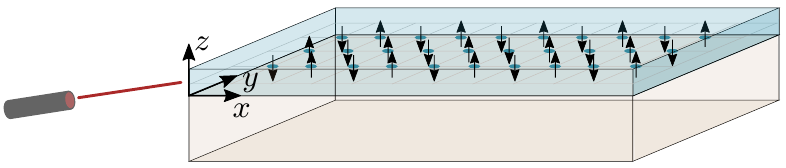}
  \caption{Sketch of an antiferromagnetically ordered strongly correlated film (top layer, with spins indicated) driven by a radiation field and in contact with a metallic reservoir (bottom layer) kept at thermal equilibrium.}
\label{fig:Fig1}
\end{figure}

\textit{Model.---}
We consider the situation sketched in  Fig.~\ref{fig:Fig1}: the half-filled two-dimensional square-lattice Hubbard model with nearest-neighbor hopping, repulsive interaction, brought out of equilibrium via an applied electromagnetic field, and tunnel-coupled to a metallic reservoir to allow the system to reach a non-equilibrium steady state. The Hamiltonian is
\begin{equation}
\hat{H}=\sum_{\bm{k}\sigma}\epsilon_{\bm{k}}^{\phantom\dagger\!}(t)\hat{c}^\dagger_{\bm{k}\sigma}\hat{c}_{\bm{k}\sigma}^{\phantom\dagger}+U\sum_{i}\h{n}_{i\uparrow}\h{n}_{i\downarrow}
+\h{H}_\text{{res}},
\label{H}
\end{equation}
where $\epsilon_{\bm{k}}(t)$ is the electron dispersion and $U$ the on-site repulsion. The operator $\hat{c}^\dagger_{i\sigma}$ creates an electron of spin $\sigma$ at site $i$ of a two-dimensional lattice of unit lattice constant, $c^\dagger_{\bm{k}\sigma}$ is its Fourier transform in the first Brillouin zone, and $\h{n}_{i\sigma}=\hat{c}^\dagger_{i\sigma}\hat{c}^{\phantom\dagger\!}_{i\sigma}$. $\h{H}_\text{res}$ is a weak tunnel coupling to an infinite-bandwidth reservoir with flat density of states \footnote{For details on the coupling to an infinite-bandwidth flat-band reservoir and how it leads to broadening, see e.g. \cite{Takei:Dissipation}} giving rise to a constant inverse electron lifetime, $\Gamma$ (see Eq.~\eqref{eq:GRA} below). We set the chemical potential corresponding to half filling, set $\hbar=k_{B}=e=1$, and include the electric field via the Peierls substitution with vector potential $A_{x,y}(t)=-E\sin(\Omega t)/\Omega$:
\begin{equation}\label{eq:PP}
\epsilon_{\bm{k}}(t)=-2\tilde{t}\left\{\cos[k_x+A_x(t)]+\cos[k_y+A_y(t)]\right\}.
\end{equation}
Henceforth, all energies are given in units of the nearest-neighbor-hopping matrix element $\tilde{t}$.

The equilibrium properties of the model are well understood~\cite{Schrieffer:Spin,Schrieffer:Dynamic,Singh:Spin}: The ground state is antiferromagnetically (N\'eel) ordered, has a gap to electronic excitations and supports gapless spin waves. The thermal population of magnons diverges as their energy goes to zero, which in turn leads to the destruction of long-ranged magnetic order at any non-zero temperature in dimension $d\leq 2$~\cite{Hohenberg:LRO,Mermin:Absence,Auerbach:Interacting}. These features are revealed by an appropriate interpretation of the results of a conventional mean field plus fluctuation analysis~\cite{Schrieffer:Spin,Schrieffer:Dynamic}, which is known to provide a qualitatively correct description of the equilibrium properties of the model.

We study the model for drive frequencies ranging from  smaller than the  equilibrium gap  (``sub-gap drive regime") to larger than the highest electronic  transition visible in linear response (``Magnus drive regime")~\cite{Magnus:On,Bukov:Universal,Mentink:Ultrafast,Kennes:Floqueteng} by solving  the non-equilibrium mean-field equations in the presence of the periodic drive  and then computing one-loop corrections. 

\textit{Saddle point approximation.---}
To generate the mean-field theory we write the model as a Keldysh-contour path-integral \cite{Kamenev:Field}, decouple the interaction via a magnetic-channel Hubbard-Stratonovich field~\cite{Coleman:Introduction}, $\bm{m}$, and consider $\bm{m}$ to have a mean-field part, $m_0\hat{\bm{z}}e^{i\bm{Q}\cdot\bm{R}_i}$, identified with the N\'eel order parameter, $\bm{Q}=(\pi,\pi)$, and a fluctuation part, $\delta \bm{m}$, which when treated to one-loop order reveals the spin-wave physics.

\begin{figure}[t!]
  \centering
  \includegraphics[width=1\columnwidth]{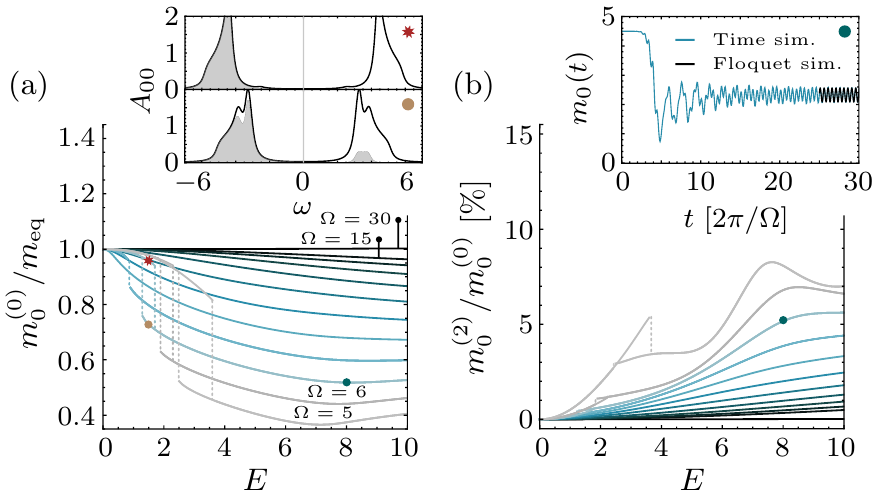}
  \caption{Mean-field solutions for varying drive frequencies $\Omega=5-15$ in steps of 1 as well as $\Omega=30$. (a) Time-averaged mean field as a function of drive amplitude, and (inset) diagonal-component of the time-averaged spectral functions (solid lines) and occupation functions (shaded areas) for the mean-field solutions marked in (a). (b) Second mean-field Floquet component as a function of drive amplitude, and (inset) an explicit time-dependent mean-field solution for $\Omega=7$ (see Refs.~\cite{Kennes12_t1,Kennes12_t2} for computational details) ramped from the un-driven, to the driven state synchronized to the time-transformed Floquet mean-field solution. The parameters are $I=5$, $T=0.01$, $\Gamma=0.2$ and $n_{\text{max}}=10$.}
\label{fig:Fig2}
\end{figure}

In a non-equilibrium steady state the mean field is synchronized to the drive  (see inset Fig. \ref{fig:Fig2}(b)) so the mean-field magnetization can be represented as a Fourier series $m_0(t)=\sum_nm^{(n)}_0e^{-in\Omega t}$. The mean-field equation, found as a saddle-point approximation for the classical magnetization field component \cite{Kamenev:Keldysh}, is then a nonlinear equation for the components $m_0^{(n)}$ of the Floquet-space vector representing $m_0(t)$
\begin{equation}\label{eq:MF}
m^{(n)}_0\!=\frac{I}{4\pi N i}\sum_{\bm{k}}{}^{\!'}
\!\int_{-\infty}^{\infty}\!\!\!\!\!\!d\omega\,
\text{Tr}\!\left[\h{\mathcal{G}}_{\bm{k},n0}(\omega)
(\h{\tau}_1\!\otimes\!\tau_1\!\otimes\!\sigma_3)\right]\!,
\end{equation}
where the primed sum is taken over the magnetic Brillouin zone (BZ), i.e. half of the electronic BZ, $I= U/3$ \cite{Coleman:Introduction}, and $\h{\mathcal{G}}$, the mean-field Floquet Green function \cite{Kitagawa:Transport,Aoki:Nonequilibrium,Eissing2016b,Kennes2018c}, is a matrix in Keldysh ($\h{\tau}$), momentum-spinor ($\tau$), spin ($\sigma$), and Floquet space. The retarded/advanced component of the electron Green's function dressed by the reservoir is given by
\begin{equation}\label{eq:GRA}
\mathcal{G}_{\bm{k},mn}^{R/A\, -1}(\omega)=(\omega+n\Omega\pm i\Gamma)\delta_{mn}\tau_0\otimes\sigma_0-h_{\bm{k},mn},
\end{equation}
where $h_{\bm{k},mn}=\epsilon_{\bm{k},m-n}\tau_3\otimes\sigma_0-m_0^{(m-n)}\tau_1\otimes \sigma_3$, with $
\epsilon_{\bm{k},m}=\frac{1}{T}\int_{-T/2}^{T/2}dt\,e^{im\Omega t}\epsilon_{\bm{k}}(t)$, describing electrons driven by the external field and moving in a time-periodic magnetization field.
The Keldysh Green's function is given by $
\mathcal{G}_{\bm{k},mn}^K(\omega)=\sum_{m'n'}
\mathcal{G}^R_{\bm{k},mm'}(\omega)\Sigma_{\bm{k},m'n'}^K(\omega)\mathcal{G}_{\bm{k},n'n}^A(\omega),$ where $\Sigma_{\bm{k},mn}^{K}(\omega)=-2i\Gamma \tanh((\omega+n\Omega)/2T)\tau_0\otimes\sigma_0\delta_{mn}$ is the self-energy from coupling to the reservoir.
We solve Eqs.~\eqref{eq:MF} and \eqref{eq:GRA} numerically, choosing a Floquet cutoff $|n|\leq n_{\text{max}}$, and iterate from an initial guess $m_0^{(n)}=10^{-2}\theta(n_{max}-|n|)$. We use converged solutions as new starting points to explore multistability.

Representative results for the zeroth Floquet component, corresponding to the time-averaged dynamics, are shown in the left-hand panel of Fig.~\ref{fig:Fig2}. For $I\gg \tilde{t}$ the qualitative physics does not depend on the interaction strength, so we present results only for a single typical case.   In the high-frequency ('Magnus') limit, $\Omega\gg2m_0^{(0)}$,  theoretical arguments~\cite{Magnus:On} suggest that the system is described by an effective Hamiltonian with hopping amplitude modified from the equilibrium value. We see that indeed on the mean-field level, the main features of the solution remain similar to equilibrium but with parameters renormalized as expected: a magnetic insulating state with the expected~\cite{Magnus:On} small increase in the average staggered magnetization (barely  visible in the $\Omega=30$ trace in panel (a) of  Fig.~\ref{fig:Fig2}) arising from the Magnus-regime renormalization of $\tilde{t}$  by $J_0(E/\Omega)$~\cite{Mentink:Ultrafast,Kennes:Floqueteng}. However, as the drive frequencies are decreased towards the sub-gap regime (drive frequency within or below the region of particle-hole continuum excitations) we observe a change to a weak decrease of the order parameter with drive amplitude, and for still lower drive frequency the mean-field equation gives a discontinuous transition (within a regime of bistability) to a state of lower gap amplitude and significant occupation of the upper band (Fig.~\ref{fig:Fig2}(a) inset). Within our mean-field theory, the state remains magnetically ordered on both sides of the transition; whether a more sophisticated approximation  as in Ref. \cite{Jose:Ultra} would lead to a Mott or gapless state is an important open question.

Figure~\ref{fig:Fig2}(b) presents the harmonic content of the order parameter. The spin inversion symmetry of the drive implies that only even harmonics of the drive frequency appear in the order parameter, and we find generically that only the 0 and $\pm 2$ Floquet components have appreciable amplitudes.
The resulting $2\Omega$ oscillation in the order parameter implies moderate second harmonic amplitude oscillations in the gap magnitudes (see inset Fig.~\ref{fig:Fig2}(b)); the resulting nonlinear optical effects will be strongest for incident radiation at frequencies near the gap.

\textit{Fluctuations.---}
We now focus on the mean-field solutions at higher drive frequency, where the density of electron quasiparticle excitations is negligible.
We introduce the  fluctuation field as a Keldysh and momentum-spinor, $\delta \bm{m}_{\bm{q}}^{\mu,i}(t)=(\delta m^{\mu,i}_{\bm{q}}(t),\delta m^{\mu,i}_{\bm{q}+\bm{Q}}(t))$ with Keldysh index $i=c,q$ (classical, quantum \cite{Kamenev:Field}) and $\mu=\pm$ referring to the directional polar decomposition $x\pm iy$. The fluctuations are governed by the electron Green function bubble, which upon transforming to Floquet space reads
\begin{align}
\Pi^{\mu\nu,ij}_{0/Q,\bm{q},mn}(\omega)=&
\frac{i}{2N}\sum_{\bm{k}}{}^{\!'}
\!\sum_{m'}\!\int_{-\infty}^{\infty}\!\frac{d\omega'}{2\pi}\text{Tr}\left[
(\h{\gamma}_{i}\!\otimes\!\tau_0\!\otimes\!\sigma_\mu)\right.\nonumber\\
&\hspace*{-25mm}\left.\times\h{\mathcal{G}}_{\bm{k},mm'}(\omega')
(\h{\gamma}_{j}\!\otimes\!\tau_{0/1}\!\otimes\!\sigma_\nu)
\h{\mathcal{G}}_{\bm{k}+\bm{q},m'n}(\omega'\!-\!\omega\!-\!n\Omega)
\right]\!,
\end{align}
with Keldysh indices encoded in the matrices $\h{\gamma}_{c/q}=\h{\tau}_{0/1}$~\cite{Kamenev:Keldysh}. Using the sublattice matrix structure~\cite{Schrieffer:Dynamic}
\begin{align}
\bm{\Pi}_{\bm{q}}=
\!\left(
\begin{array}{cc}
\Pi_{0,\bm{q}} & \Pi_{Q,\bm{q}}\\
\Pi_{Q,\bm{q}} & \Pi_{0,\bm{q}+\bm{Q}}
\end{array}
\!\right)\!,
\end{align}
we define the corresponding transverse fluctuation matrix propagator, $\bm{\chi}^{\perp,ij}_{\bm{q}}(t,t')=(iN/\pi)\langle \delta m^{+,i}_{\bm{q}}(t)\delta m^{-,j}_{-\bm{q}}(t')\rangle$, as
\begin{align}
\bm{\chi}^{\perp R/A}_{\bm{q},mn}\!=&
\!\left[(2I)^{-1}\!-\bm{\Pi}_{\bm{q}}^{\perp R/A}\right]^{-1}_{mn}
,\\
\bm{\chi}^{\perp K}_{\bm{q},mn}\!=&
\!\left[(2I)^{-1}\!-\bm{\Pi}_{\bm{q}}^{\perp R}\right]^{-1}_{mm'}\!\bm{\Pi}_{\bm{q},m'n'}^{\perp K}
\!\!\left[(2I)^{-1}\!-\bm{\Pi}_{\bm{q}}^{\perp A}\right]^{-1}_{n'n}\!.\nonumber
\end{align}

\begin{figure}[t!]
  \centering
  \includegraphics[width=1\columnwidth]{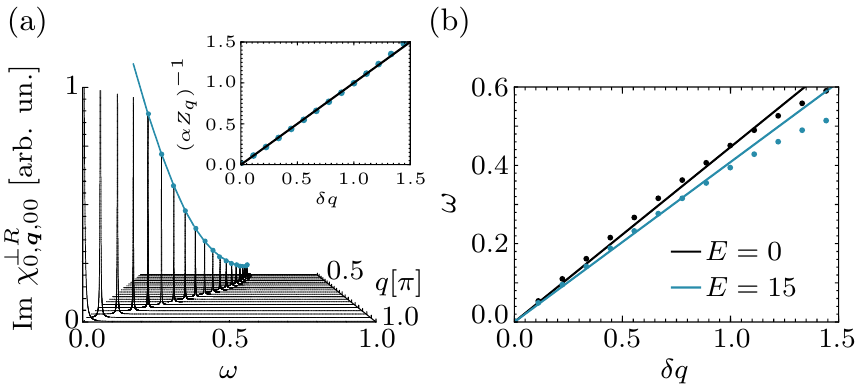}
  \caption{Transverse spin wave modes. (a) Imaginary part of the retarded susceptibility as function of frequency and momentum $q_x=q_y=q$ for $E=15$ showing the spin wave pole. Inset: Inverse spectral weight of the peaks in (a). (b) Location of the spin wave pole (points) as a function of frequency and $\delta q$ together with the equilibrium linear spin wave dispersion, $\omega=v\delta q$, (solid lines) with $\tilde{t}\to\tilde{t}J_0(E/\Omega)$. The parameters are $I=5$, $\Omega=30$, $T=0.01$, $\Gamma=0.2$, and $n_{\text{max}}=3$.}
\label{fig:Fig3}
\end{figure}

The time-averaged ($00$-Floquet) fluctuation spectrum is revealed by ${\rm Im}\chi_{\bm{0},\bm{q},00}^{\perp,R}(\omega)$, shown in  the left panel of Fig.~\ref{fig:Fig3}. We see that the only low-lying excitations are very sharp peaks, corresponding to spin waves, with a small but non-zero broadening from the coupling to the reservoir. The peak energy vanishes  and the peak amplitude grows as $\bm{q}\rightarrow \bm{Q}$. At energies below the charge gap, for positive frequencies ${\rm Im}\chi^{\perp R}_{\bm{0},\bm{q},00}(\omega)\approx Z_q\delta(\omega-\omega_q)$ for not too large $\Gamma$. Upon integrating over the peaks in Fig.~\ref{fig:Fig3}(a), the inverse spectral weight $Z_q^{-1}$ shows a linear $\delta q=|\bm{q}-\bm{Q}|$ dependence (Fig. \ref{fig:Fig3}(a) inset) which agrees well with the expanded equilibrium result, $Z_q^{-1}\approx\alpha\delta q,\ \alpha=1/(8\sqrt{2}\pi m_{0}^{2})[2+t^{2}/m_{0}^{2}+\mathcal{O}(t^{4}/m_{0}^{4})]$.
The $\omega_q$ is determined from the peak positions, and gives the dispersions presented in the right panel of Fig.~\ref{fig:Fig3}. The dispersion exhibits the expected linear momentum dependence at lowest energies, $\omega=v\delta q$. The spin wave velocity is seen to compare  well to the dissipative equilibrium result, $v=(2\sqrt{2}\tilde{t}^2/m_0)(1-5\tilde{t}^2/m_0^2-3\Gamma/\pi m_0-\Gamma^2/2m_0^2)+\mathcal{O}\left(\tilde{t}^{2+n}\Gamma^{3-n}/m_0^5\right)$ for $n=0,1,2,3$ (consistent with Ref.~\cite{Singh:Spin} for $\Gamma=0$), provided that the hopping amplitude $\tilde{t}$ is replaced by the Magnus-renormalized value  $\tilde{t}J_0(E/\Omega)$~\cite{Mentink:Ultrafast}.
One may view this Bessel-function reduction of the spin-wave velocity as a particularly simple example of "Floquet engineering".

The Keldysh component of the transverse propagator contains information about the non-equilibrium distribution of excitations. For low-lying spin waves with $\omega_{q}\ll\Omega$, this information resides in the zeroth Floquet component, from which we define a time-averaged distribution function, $F$, by the ansatz
\begin{equation}
\begin{split}
\chi^{\perp K}_{\bm{0},\bm{q},00}(\omega)&=
2i\,{\rm Im}\!\left[\chi^{\perp R}_{\bm{0},\bm{q},00}(\omega)\right]\!F(\bm{q},\omega)\\
&\approx 2iZ_q\delta(|\omega|-\omega_q)F_q.
\end{split}
\label{chiK}
\end{equation}
The spin-wave pole approximation to Im$\chi^R$ allows for a quasiclassical description in terms of an on-shell distribution function, $F_q=F(\bm{q},\omega_q)$, referring only to the mode energy $\omega_q$. In equilibrium, the fluctuation-dissipation theorem (FDT) ensures that $F_q=\coth(\omega_q/2T)$, which tends to unity at $\omega_q\gg T$ and diverges as $\omega_q^{-1}$ for $\omega_q\to 0$.

\begin{figure}[t!]
  \centering
  \includegraphics[width=1\columnwidth]{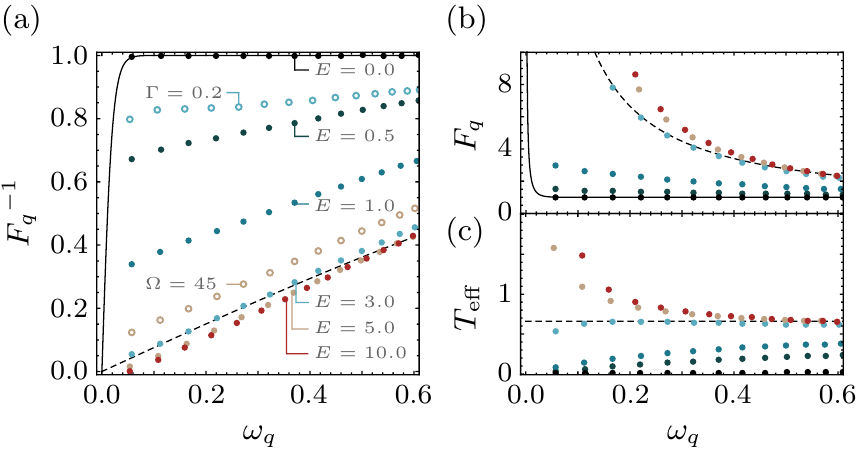}
  \caption{(a) $F_{q}^{-1}$ and (b) $F_{q}$ as function of $\omega_q$ for increasing drive amplitude with $\Gamma=0.02$ and $\Omega=30$ together with the equilibrium curves for $T=0.01$ (solid) and $T=0.66$ (dashed). In (a) is also shown the result for $\Omega=45$, $E=5.0$, $\Gamma=0.02$ and $\Gamma=0.2$, $E=3.0$, $\Omega=30$. (c) $T_{\text{eff}}$ corresponding to the curves in (b) together with the equilibrium $T=0.66$ line (dashed). The parameters are $I=5$, $T=0.01$, and $n_{\text{max}}=3$.}
\label{fig:Fig4}
\end{figure}

Figure~\ref{fig:Fig4}(a) shows the inverse distribution function, $F_q^{-1}$, as a function of the mode energy, $\omega_q$, at different drive amplitudes for a low reservoir temperature, $T=0.01$. We plot the reciprocal to fit all data on the same panel. Because the reservoir temperature is substantially lower than the lowest $\omega_q$ included in our numerics, the equilibrium $F_q$ (Fig.~\ref{fig:Fig4}(b)) is indistinguishable from unity. We see that increasing the drive amplitude increases $F_q$  (decreases $F_q^{-1}$) at all $\omega_q$, with a larger increase for lower $\omega_q$. Increasing either the drive frequency, $\Omega$, or the reservoir coupling, $\Gamma$, for fixed drive amplitude  reduces $F_q$ (open symbols, left panel Fig.~\ref{fig:Fig4}). For higher $\omega_q$, $F_q$ initially increases rapidly as the drive amplitude increases, but then saturates as the amplitude becomes large. For small $\omega_q$, the situation is different. For the two weakest drive amplitudes, $F_q$ appears to approach a finite, non-zero value as $\omega_q$ approaches zero; for the intermediate drive amplitude $F_q^{-1}$  vanishes linearly as $\omega_q\to 0$ while for the two highest drive amplitudes, $F_q^{-1}$ vanishes faster than linearly as $\omega_q\rightarrow 0$.

Apart from the intermediate drive amplitude ($E=3$), these  distribution functions depart markedly from the equilibrium distribution dictated by the FDT. To illustrate this more clearly, Fig.~\ref{fig:Fig4}(c) shows the effective temperature $T_{\text{eff}}$ as defined by $F_q=\coth(\omega_q/2T_{\rm eff}(q))$. We see that the results fall into two groups. For the two smallest drive amplitudes, $T_{\rm eff}$ is larger at high $\omega_q$ (very substantial excitation of high $q$ spin waves above the equilibrium value), but decreases to a value consistent with the reservoir temperature as $\omega_q\to 0$. For the intermediate drive amplitude, $T_{\rm eff}\approx 0.66$ is essentially momentum-independent (i.e. $F_q$ fits well to the equilibrium form) and much larger than the reservoir temperatures. For the two larger drive amplitudes, $T_{\rm eff}$ increases rapidly for small $\omega_q$, indicating a super-thermal occupancy of the low-lying spin wave modes, in other words $F_q$ diverging faster than $1/\omega_q$. 

The site- and period-averaged mean squared fluctuations of the classical component of the order parameter are given by
\begin{equation}\label{eq:fluct}
\langle|\delta m^{+,c}|^2\rangle=\!\frac{1}{N}\!\sum_{\bm{q}}
\!\int\!\!\frac{d\omega}{4\pi i}\chi_{0,\bm{q},00}^{\perp K}(\omega) \sim\!\int\!\!\frac{d^2q}{(2\pi)^2}Z_q F_q.
\end{equation}
In thermal equilibrium at any non-zero temperature, both $F_q$ and $Z_q$ diverge as $1/\delta q$, and $\langle|\delta m^{+,c}|^2\rangle$ therefore diverges logarithmically with system size in two dimensions. This is the expression in the one-loop calculation of the well-known  result~\cite{Hohenberg:LRO,Mermin:Absence} that thermal fluctuations destabilize long-ranged magnetic order in continuous-symmetry systems of dimension $d\leq2$.  Our results indicate that the generalization to systems out of equilibrium is richer than expected from previous work.  Unlike the dc current-driven ferromagnetic case~\cite{Mitra:Nonequilibrium,Mitra:Current}, a weak non-equilibrium drive would not destabilize the ordered state for $d=2$, but larger drives lead to a super-thermal occupancy that can  destabilize the order even in $d>2$.

\textit{Conclusions.---}
We have used a mean field plus fluctuation analysis of the antiferromagnetic two-dimensional Hubbard model driven by an oscillating electric field to examine the   accepted theoretical intuition, which suggests that if an ac drive is detuned from direct electronic transition energies, its main effect is to renormalize Hamiltonian parameters.  Our solution of the full non-equilibrium problem shows rich additional physics: i) in the sub-gap drive regime, the drive is found to induce a substantial time-dependent component of the order parameter with first-order like transitions and coexistence regimes involving several locally stable (at least at the mean-field level) phases, and ii) in all cases, including the  ``Magnus'' regime of very high frequency drive where the basic electronic state evolves smoothly with drive amplitude and no electronic quasiparticle excitations are created, we find a highly non-thermal distribution of magnons. Whereas the main focus in this paper is on the latter, an analysis of fluctuation effects on the bistability observed in the sub-gap drive regime is an interesting open question.

The interaction-mediated transfer of energy to the spin fluctuations may be thought of as a spin-charge coupling (albeit a weaker kind than considered e.g. in Ref. \cite{Dolcini:Signature}). The dependence of the magnon distribution on the drive frequency and coupling to the reservoir indicates that the pathway to spin wave excitation involves reservoir states. The kinetics of this process, and the generalization to more realistic models of solids, are an important subject for future research. The distribution of fluctuations  depends in a remarkable way on the drive amplitude.  For small and moderate drive amplitude, there is substantial excitation of higher energy modes, but as the momentum tends to the ordering wave vector, the distribution tends towards the equilibrium one. However, at larger drive amplitude, the distribution diverges faster than $\omega_q^{-1}$ as momentum tends towards the ordering wave vector, which would indicate destabilization of order even in three dimensions.  This apparent dynamical phase transition as a function of drive amplitude requires further study.

More generally, our findings show that the low-lying collective degrees of freedom are generically  excited by the drive, and have a large, typically non-thermal, and drive amplitude-dependent  occupancy that can lead to remarkable effects on physical properties. This finding calls into question the Floquet engineering paradigm in which applied radiation changes the Hamiltonian without changing the distribution function.

\textit{Acknowledgements.---} The Center for Nanostructured Graphene (Proj. DNRF103) and the Center for Quantum Devices are sponsored by the Danish National Research Foundation. A.J.M. and D.M.K. were supported by the US Department of Energy, Office of Basic Energy Sciences, Division of
Materials Sciences and Engineering Grant DE SC0018218. D.M.K. additionally acknowledges support by the Deutsche Forschungsgemeinschaft through the Emmy Noether program (KA 3360/2-1). N.W. thanks Antti-Pekka Jauho for useful discussions.

%

\end{document}